\pdfoutput=1
\documentclass[aps,prx,reprint,superscriptaddress,longbibliography,nofootinbib]{revtex4-2}
\usepackage{graphicx}
\usepackage{amsmath}
\usepackage{amssymb}
\usepackage{amsfonts}
\usepackage{dcolumn}
\usepackage{dsfont}
\usepackage{latexsym}
\usepackage{rotating}
\usepackage{color}
\usepackage{latexsym}
\usepackage{bbm}
\usepackage{subfigure}
\usepackage{float}
\usepackage{epsfig}
\usepackage{psfrag}
\usepackage{natbib, hyperref}
\usepackage{bm}
\usepackage{amsthm}
\usepackage{siunitx}
\usepackage{eucal}
\usepackage{mathrsfs}
\usepackage{url}
\usepackage{braket}
\usepackage{ulem}
\usepackage{calligra}
\usepackage[utf8]{inputenc}
\usepackage{newunicodechar}
\usepackage{marvosym}
\usepackage[absolute]{textpos}
\usepackage[cal=cm]{mathalfa}
\usepackage{soul}
\usepackage{lipsum}
\usepackage{upgreek} 
\usepackage{float}

\usepackage{color}

\usepackage{hyperref}
\hypersetup{
colorlinks=true,final=true,
        linkcolor=blue,
        citecolor=blue,
        filecolor=blue,
        urlcolor=blue,
}

\begin{document}
\setlength{\abovedisplayskip}{3pt}
\setlength{\belowdisplayskip}{3pt}

\title{Magnetic field-free braiding and nontrivial fusion of Majorana bound states\\ in a high-temperature planar Josephson junction}
\author{Pankaj Sharma}
\email[]{pankaj@ph.iitr.ac.in}
\affiliation{Department of Physics, Indian Institute of Technology Roorkee, Roorkee 247667, India}
\author{Narayan Mohanta}
\email[]{narayan.mohanta@ph.iitr.ac.in}
\affiliation{Department of Physics, Indian Institute of Technology Roorkee, Roorkee 247667, India}


\begin{abstract}
Demonstration of non-Abelian statistics of Majorana bound states (MBS) is crucial for the realization of fault-tolerant topological quantum computation. Two-dimensional platforms such as planar Josephson junctions require an in-plane magnetic field to generate a pair of MBS at its non-superconducting channel ends; however, the fixed direction of the in-plane magnetic field puts a constraint on the realization of a multi-terminal topological planar junction, and hence its ability to physically move multiple MBS---which is necessary for performing the fusion and braiding operations. Here we show that in a planar Josephson junction coupled to a skyrmion crystal, which can generate multiple pairs of MBS in the absence of any external magnetic field, the non-trivial fusion and braiding operations can be performed. Our numerical calculations, designed for realistic two-dimensional quantum systems, certify the feasibility of experimental realization of the proposed device schemes. We find that both $s$-wave and $d$-wave superconducting leads can generate the MBS; indicating that the MBS movement operations can be performed at higher temperatures using $d$-wave superconducting leads. Our results establish that the skyrmion crystal-coupled planar Josephson junction is a viable platform for the generation and controlled movement of the MBS.
\end{abstract}

\maketitle

\section{Introduction}
\label{sec1}
\vspace{-2mm}
Majorana bound states (MBS), the elusive quasi-particles that are known to be their own anti-particles, are at the hotspot of research in physics and computer science primarily for their application in topological quantum computation since their non-Abelian statistics can make the computation free from qubit decoherence~\cite{Kitaev_2000, Ivanov_2001, Kitaev_2003, Nayak_RMP2008, Sau_PRL2010, Fu_PRL2008, Aguado_PRL2012, Oreg_PRL2010, Alicea_Nature2011, Alicea_2012, Beenaker_SciPost2020}. In recent years, extensive efforts have been made to detect the MBS using various signatures, including the quantized zero bias conductance peak and the 4$\pi$-periodic Josephson effect in different platforms; however, the identification of the MBS appears to be more non-trivial than estimated, due to the ubiquitous presence of Andreev bound states and disorder-induced localized states~\cite{Schiela_PRXQuantum2024}. Hence, it is believed that the existence of the MBS in a topological superconductor should be confirmed via experimental demonstration of their non-Abelian exchange statistics, and for this purpose a two-dimensional platform, in which multiple MBS can be moved physically, is suitable. A planar Josephson junction (PJJ) based on a semiconductor/superconductor heterostructure stands out from other proposed geometries, as it has advantages such as intrinsic stability due to two-dimensional geometry, tunability with respect to magnetic field, gate voltage and phase bias~\cite{Hell_PRL2017,Pientka_PRX2017,Fornieri_Nature2019,Ren_Nature2019,Stern_PRL2019,Nichele_PRL2020,Laeven_PRL2020,Mohanta_CommunPhys2021,Moehle_NanoLett2021,Peng_PRR2021,Dartiailh2021,Haxell_PRL2023,Banerjee_PRL2023,Banerjee_PRB2023,Vakili_PRB2023,Kuiri_PRB2023,Leuthi_PRB2023,Xie_PRL2023,Sharma_PRB2024,Schiela_PRXQuantum2024,Subramanyan_PRB2024,Sharma_arXiv2024}. Despite these advantages, the standard PJJ model lacks scalability, required to realize topological superconductivity in a multichannel PJJ geometry, due to the fixed direction of the external in-plane magnetic field.

In this work, we show that non-trivial fusion and non-Abelian braiding operations of the MBS can be performed in the absence of any external magnetic field in PJJ devices coupled to a  skyrmion crystal (SkX). The SkX provides, to the two-dimensional electron gas (2DEG) of the PJJ by proximity effect, a spatially-varying local magnetic field that generates the necessary Zeeman splitting of the energy levels, and also a gauge field similar to the Rashba spin-orbit coupling. As a result, a topological superconductivity is induced in the entire proximitized 2DEG within a suitable range of doping, when the phase difference between the two superconducting leads is zero. The non-superconducting middle channel of the PJJ acts as a line defect; and when the PJJ is topological, it hosts a MBS pair at its two ends. 

There has been a huge progress in generating chiral magnetism such as a SkX with nanoscale control via advanced interface engineering~\cite{Jiang2017,Cheng_PRB2023,Sun_PRL2013}. Combining a non-trivial magnetic texture with superconductivity to generate topological superconductivity, therefore, opens a promising route to build complex magnetic field-free architectures for controlled manipulations of the MBS, and relaxes the need of a strong Rashba-type spin-orbit coupling to obtain the desired topological pairing symmetry~\cite{Mohanta_PRApp2019,Mohanta_CommunPhys2021,Daniel_2022}.

The advantages of this SkX-coupled PJJ over the conventional PJJ are the following: (i) multi-terminal topological PJJs, that can host multiple pairs of the MBS, can be realized, as we demonstrate below numerically, (ii) the detrimental effect of the magnetic field on the proximity-induced superconductivity in semiconductor-superconductor heterostructures poses challenges for the realization of topological superconductivity; this issue can be resolved in a field-free platform such as the one presented here, (iii) both $s$-wave and $d$-wave superconductors used as the leads in our SkX-coupled PJJ generate the MBS, implying the flexibility of using different superconducting materials and working with the MBS at higher temperatures, (iv) the skyrmion radius, when controlled externally, brings an additional knob to turn ON or OFF the MBS in the PJJ platform, (v) the MBS are obtained at zero phase difference between the two superconducting leads, eliminating the need of a $\pi$-phase biasing requirement as in the conventional PJJ set up. A SkX-coupled PJJ can, therefore, be a scalable magnetic field-free platform that can help to achieve experimental demonstration of the non-Abelian statistics of the MBS and implementation of fault-tolerant quantum gates. 

The rest of the paper is organized as follows. In Sec.~\ref{sec2}, we describe our theoretical model and the method of calculation. In Sec.~\ref{sec3}, we numerically demonstrate non-trivial fusion of the MBS using a barrier potential. In Sec.~\ref{sec4}, we numerically demonstrate exchange of two MBS in a T-shape PJJ geometry. In Sec.~\ref{sec5}, we discuss non-Abelian braiding procedure involving three MBS in a double-cross PJJ geometry. In Sec.~\ref{sec6}, we conclude our findings and discuss future directions of device implementation towards universal quantum computation using the SkX-coupled PJJ architectures.

\section{Topological Josephson junction with a sKyrmion crystal } 
\label{sec2}
\begin{figure*}[t]
\begin{center}
\vspace{-0mm}
\epsfig{file=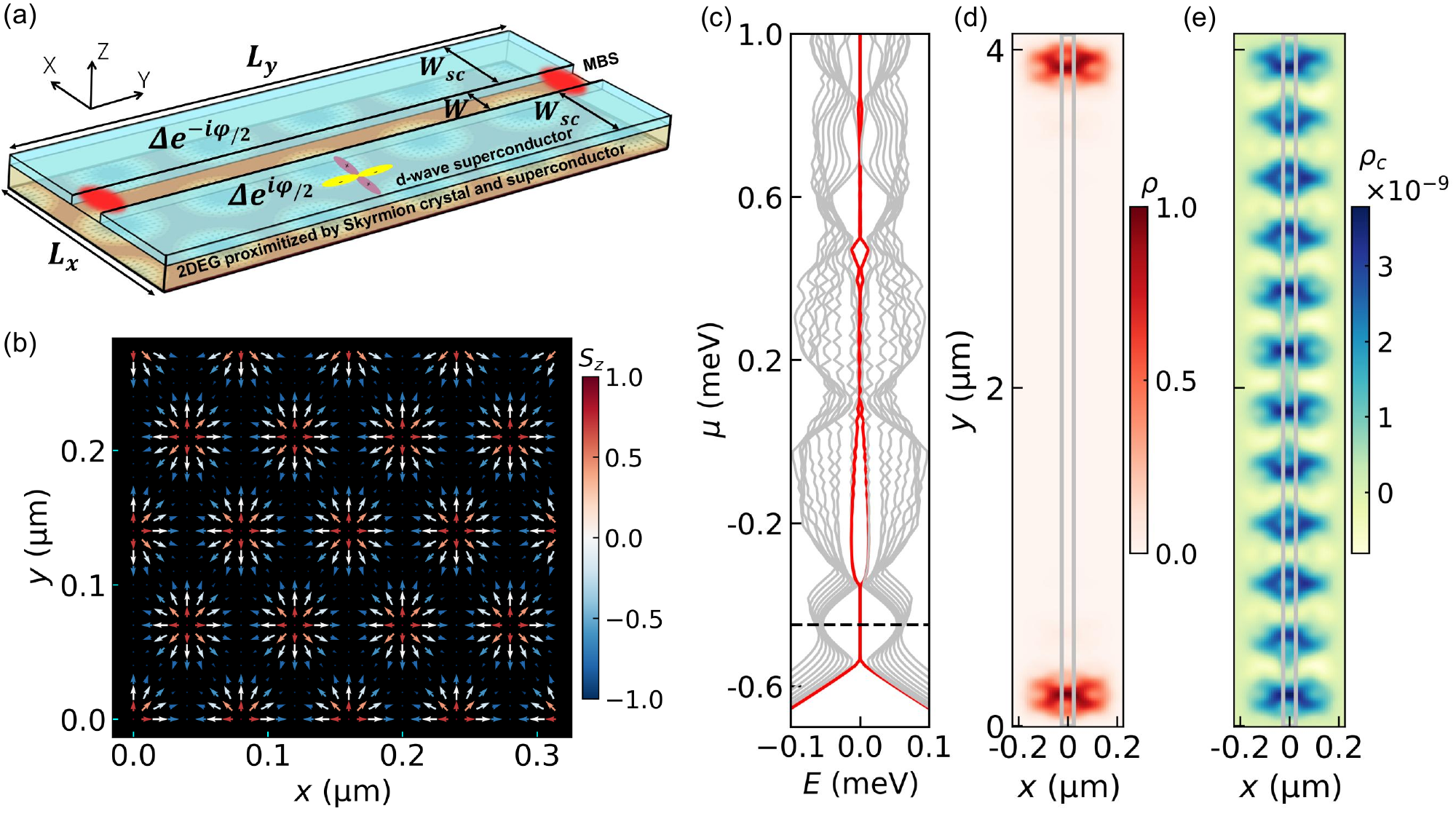,trim=0.0in 0.0in 0.0in 0.0in,clip=false, width=172mm}
\caption{(a) A planar Josephson junction with $d_{x^2 - y^2}$-wave superconducting leads and a skyrmion crystal attached underneath. The two-dimensional electron gas (2DEG) is proximitized by the superconducting regions from above and by the skyrmion crystal from below. (b) Representative spin configuration of a skyrmion crystal structure; the colorbar shows the value of the $z$-component of the spins while the arrows represent the ($x$,~$y$) components of the spins. (c) Variation of quasiparticle energy spectrum as a function of the global chemical potential $\mu$. (d) Profile of the local density of states $\rho$ (normalized to unity) at $\mu = \SI{-0.45}{\meV}$ (indicated by black dotted line in (c)), corresponding to the lowest positive energy state, revealing a pair of Majorana bound states near the ends of the middle channel. (e) Profile of the charge density of states $\rho_c$ (in units of e/$\upmu$m$^2$) at $\mu = \SI{-0.45}{\meV}$ shows a typical oscillatory pattern. The skyrmions, arranged in a triangular crystal structure, have a radius of $R_s = 60$~nm. The phase difference between the superconducting leads is kept at zero.  The dimensions of the planar Josephson junction are $L_y$ = \SI{4.1}{\micro \meter}, $L_x$ = \SI{0.45}{\micro \meter} and $W = 50$~nm.
}
\label{FIG:1}
\vspace{2mm}
\end{center}
\end{figure*}

\vspace{-2mm}
We consider a two-dimensional metallic system, such as a 2DEG in a semiconducting quantum well, which is proximitized by two superconducting lead regions from top and a SkX from below, as shown schematically in Fig.~\ref{FIG:1}(a). A SkX can be generated using a heavy metal/ferromagnet interface~\cite{Mohanta_CommunPhys2021} or by suitable interface engineering in the absence of any magnetic field~\cite{Sun_PRL2013}. We use the spin configuration of a triangular SkX, depicted in Fig.~\ref{FIG:1}(b), obtained using
\begin{align}
{\bf S}_i=S(\sin{\theta_i}\cos{\phi_i},\sin{\theta_i}\sin{\phi_i},\cos{\theta_i}),
\end{align}
where $S$ is the spin amplitude and the spin angles are defined as
\begin{align}
{\theta}_i=\pi~{\text{min}}\Big(\frac{|{\bf r}-{\bf R}_i|}{R_s},1\Big),~{\phi}_i={\text{arctan}}\Big(\frac{y-y_i}{x-x_i}\Big),
\end{align}
${\bf R}_i=(x_i,y_i)$ denote the skyrmion center near a site position ${\bf r}=(x,y)$ and $R_s$ is the skyrmion radius.

We use the below Hamiltonian to describe the 2DEG, having proximity-induced superconductivity and a coupling to the SkX field,
\begin{align}
	\mathcal{H} &= \mathcal{H}_\mu +  \mathcal{H}_{\rm{K.E.}} + \mathcal{H}_{\text{RSOC}} + \mathcal{H}_{\text{SkX}} + \mathcal{H}_\Delta,
\label{model_hamiltonian}
\end{align}
\noindent where
\begin{align*}
	&\mathcal{H}_{\mu} = \sum_{i, \sigma} (4t - \mu) c^\dagger_{i\sigma}c_{i\sigma} \\
    &\mathcal{H}_{\rm{K.E.}} = - t \sum_{\langle i j\rangle, \sigma} \left(c^\dagger_{i\sigma}c_{j\sigma} + {\rm H.c.}\right) \\
    &\mathcal{H}_{\rm{RSOC}} = - \frac{\mathbf{i}\alpha}{2a} \sum_{\langle i j\rangle, \sigma \sigma^\prime} 
	\left(\bm{\sigma} \times \mathbf{d}_{ij}\right)^z_{\sigma \sigma^\prime} c^\dagger_{i\sigma} c_{j\sigma^\prime} \\
    &\mathcal{H}_{\rm{SkX}} = \frac{g_{eff}~\upmu_{_{\rm B}} }{2} \sum_{i, \sigma, \sigma^\prime} \left(\mathbf{S}_i \cdot \bm{\sigma}\right)_{\sigma \sigma^\prime} c^\dagger_{i\sigma} c_{i\sigma^\prime} \\
     &\mathcal{H}_{\Delta} = \sum_{\langle i j\rangle} (e^{\mathbf{i}\varphi} \Delta_{ij}  c^\dagger_{j \uparrow}c^\dagger_{i\downarrow}  + {\rm H.c.}).
\end{align*}
\noindent The terms $\mathcal{H}_{\mu}$, $\mathcal{H}_{\rm{K.E.}}$, $\mathcal{H}_{\rm{RSOC}}$, $\mathcal{H}_{\rm{SkX}}$ and $\mathcal{H}_{\Delta}$ represent the doping in the 2DEG, kinetic energy of the electrons, Rashba spin-orbit coupling which arises due to broken inversion symmetry in the 2DEG, magnetic coupling of the 2DEG to the SkX texture and induced superconducting pairing, respectively. In the above equations, $t\!=\!\hbar^2/2ma^2$ is the kinetic hopping energy of electrons between nearest-neighbor sites, $m$ is the effective mass of the electrons, $a$ is the spacing of the considered square lattice grid, $i$ and $j$ represent lattice site indices, $\sigma$ and $\sigma^\prime$ are indices that represent the spin components ($\uparrow$, $\downarrow$), $\mu$ is the global chemical potential, $\Delta_{ij}$ is the nearest-neighbor $d_{x^2-y^2}$-wave superconducting pairing amplitude in the leads, $\varphi$ is the phase difference between the left and the right superconducting leads, $g_{eff}$ is the effective $g$-factor of electrons, $\upmu_{_{\rm B}}$ is the Bohr magneton, $\bm{\sigma}$ is the Pauli matrices which represent the spin of the mobile electrons in the 2DEG, $\alpha$ is the strength of the Rashba spin orbit coupling, $\mathbf{i}$ denotes the imaginary number, and $\mathbf{d}_{ij}$ is the unit vector from site $i$ to $j$.

In what follows, we diagonalize the Hamiltonian~(\ref{model_hamiltonian}) using an unitary transformation of the fermionic operators ${c}_{i \sigma} \!=\! \sum_{n}u^n_{i \sigma}{\gamma}_n + v^{n *}_{i \sigma} \gamma^\dagger_n$, where $u^n_{i \sigma}$ ($ v^{n }_{i \sigma}$) represents quasiparticle (quasihole) amplitude, and ${\gamma}_n$ (${\gamma}_n^\dagger$) is a fermionic annihilation (creation) operator. The quasiparticle (quasihole) amplitudes $u^n_{i \sigma}$ ($ v^{n }_{i \sigma}$) are obtained by solving the Bogoliubov-de Gennes equations $\sum_j {\cal H}_{ij}\psi_j^n \!=\! \epsilon_n \psi_i^n$, where ${\cal H}_{ij}$ is the matrix representation of the Hamiltonian in Eq.~(\ref{model_hamiltonian}), $\psi_i^n \!=\! [u_{i\uparrow}^n, u_{i\downarrow}^n, v_{i\uparrow}^n, v_{i\downarrow}^n]^T$ is the basis wave function and $\epsilon_n$ is the $n^{\rm th}$ energy eigenvalue. We make use of the package Kwant~\cite{Groth_NJP2014} to diagonalize the Hamiltonian matrix, and to obtain the eigenvalues and eigenvectors. The system parameters used in this study, considering InSb quantum well 2DEG as the model platform, are $m\!=\!0.017m_0$, $m_0$ being the rest mass of the electrons, $a\!=\!10$~nm, $g_{eff}\!=\!40$, $\Delta_{ij} \!=\!2$~meV and $\alpha \!=\!30$~meV-nm~\cite{Smith_PRB1987,Mayer_APL2019,Carrad_advmat2020}. The amplitude of the spin texture is considered to be $S\!=\!1$~T, which may be realized in magnetic multilayers~\cite{Zhang_JAP1991,Charilaou_PRB2016,Heigl_JAP2020,Lonsky_PRM2022}.

In Fig.~\ref{FIG:1}(c), we show the variation of the quasiparticle energy spectrum as a function of the global chemical potential $\mu$. The emergence of the zero-energy MBS occurs at $\mu \!=\! -0.54$~meV, and the MBS exist till the value of $\mu \!=\! -0.35$~meV. The profile of the local density of states (LDOS), obtained using $\rho_{i} \!=\! \sum_\sigma (|u_{i \sigma}|^2 + |v_{i \sigma}|^2)$, corresponding to the lowest-energy quasiparticle eigenstate at $\mu = -0.45~$meV is depicted in Fig.~\ref{FIG:1}(d), revealing two MBS localized near the ends of the middle channel. Fig.~\ref{FIG:1}(e) shows the profile of the charge density of states (CDOS), obtained using $\rho^c_{i} \!=\! e\sum_\sigma (|u_{i \sigma}|^2 - |v_{i \sigma}|^2)$, corresponding to the same eigenstate as above, revealing an oscillatory pattern along the channel length. These results indicate the appearance of the MBS in the SkX-coupled PJJ. In the conventional PJJ set up with an in-plane magnetic field, the MBS appear at and around a phase difference $\varphi \!=\! \pi$ between the superconducting leads~\cite{Pientka_PRX2017,Sharma_arXiv2024}. Interestingly, in our SkX-coupled PJJs, the MBS appear robustly at zero phase difference $\varphi \!=\!0$. Nonetheless, the phase difference can be used as a tuning parameter to control the existence of the MBS. Here we use $d_{x^2-y^2}$-wave superconducting leads, but $s$-wave superconducting leads also produce a robust topological superconducting phase, as shown in a previous work~\cite{Mohanta_CommunPhys2021}. Figure~\ref{FIG:1}(c) shows that there are other ranges of $\mu$ values within which the zero-energy MBS apear, similar to the PJJs in which topological superconductivity is induced by a magnetic field~\cite{Sharma_arXiv2024}.



\begin{figure*}[t]
\begin{center}
\vspace{-0mm}
\epsfig{file=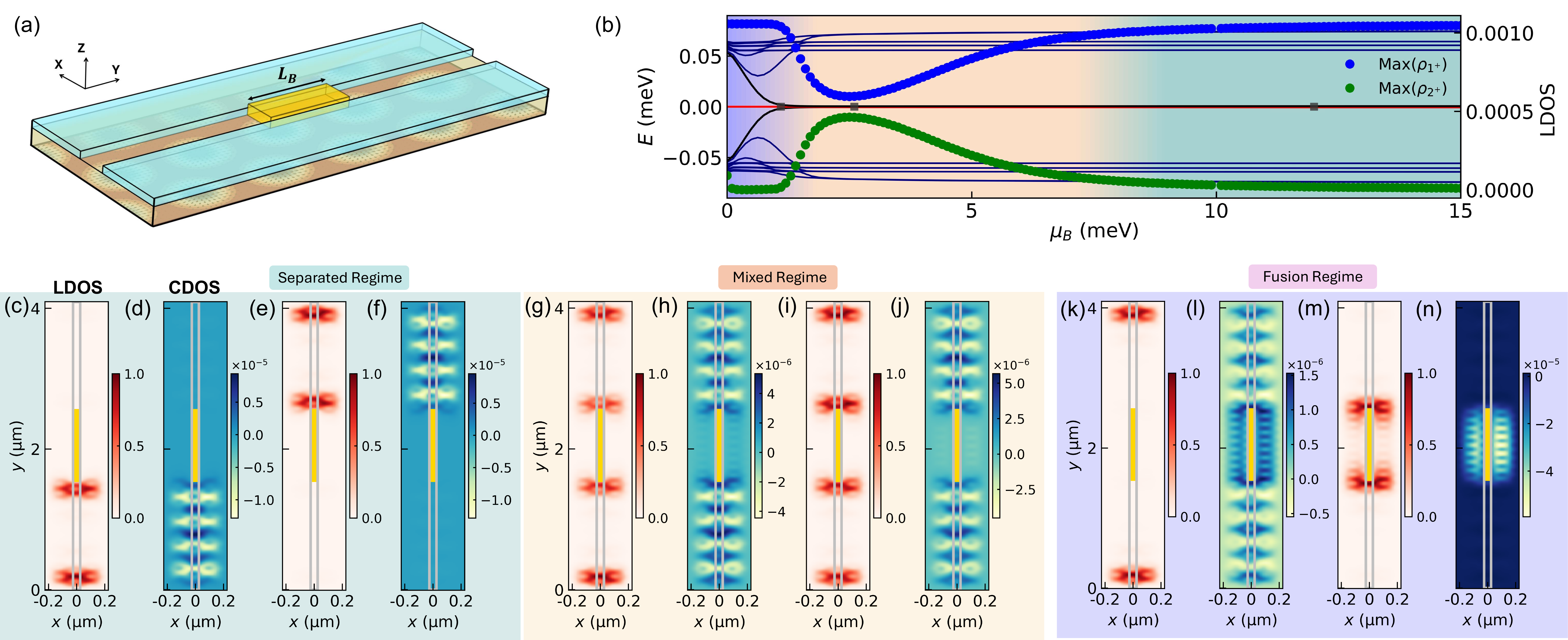,trim=0.0in 0.0in 0.0in 0.0in,clip=false, width=\textwidth}
\caption{(a) A planar Josephson junction with a potential barrier (in yellow) in the middle of the non-superconducting channel, considered for performing non-trivial fusion of two Majorana bound states (MBS). (b) Variation of the energy eigenvalues (left vertical axis) with respect to the barrier potential $\mu_B$. The global chemical potential is fixed at $\mu \!=$\SI{-0.45}{\meV}, at which the MBS appear in the junction. The right vertical axis shows the maximum value of the local density of states (LDOS) corresponding to the lowest positive and second-lowest positive energy eigenstates max$(\rho_{1^+})$ and max$(\rho_{2^+})$ respectively, in the first and second quarters of the channel, where only one Majorana bound state is localized. Based on the localization of the two pairs of the MBS, three regimes of $\mu_B$ are identified: isolated regime (in green), entangled regime (in yellow) and fusion regime (in violet). (c), (d) The profiles of the LDOS and the charge density of state (CDOS) for the lowest positive energy eigenstate. (e), (f) The LDOS and CDOS profiles for the second-lowest positive energy eigenstate at $\mu_B \!=\!12$~meV, indicated by the grey dot at zero energy in the isolated regime. (g)-(j) The same information as in (c)-(f) at $\mu_B \!=\! 2.6$~meV, indicated by the grey dot at zero energy in the entangled regime. (k)-(n) The same information as in (c)-(f) at $\mu_B \!=\! 1.1$~meV, indicated by the grey dot at zero energy in the fusion regime. The dimensions of the planar Josephson junction are the same as in FIG.1.}
\label{FIG:2}\vspace{2mm}
\end{center}
\end{figure*}

\begin{figure*}[t]
\vspace{-0mm}
\epsfig{file=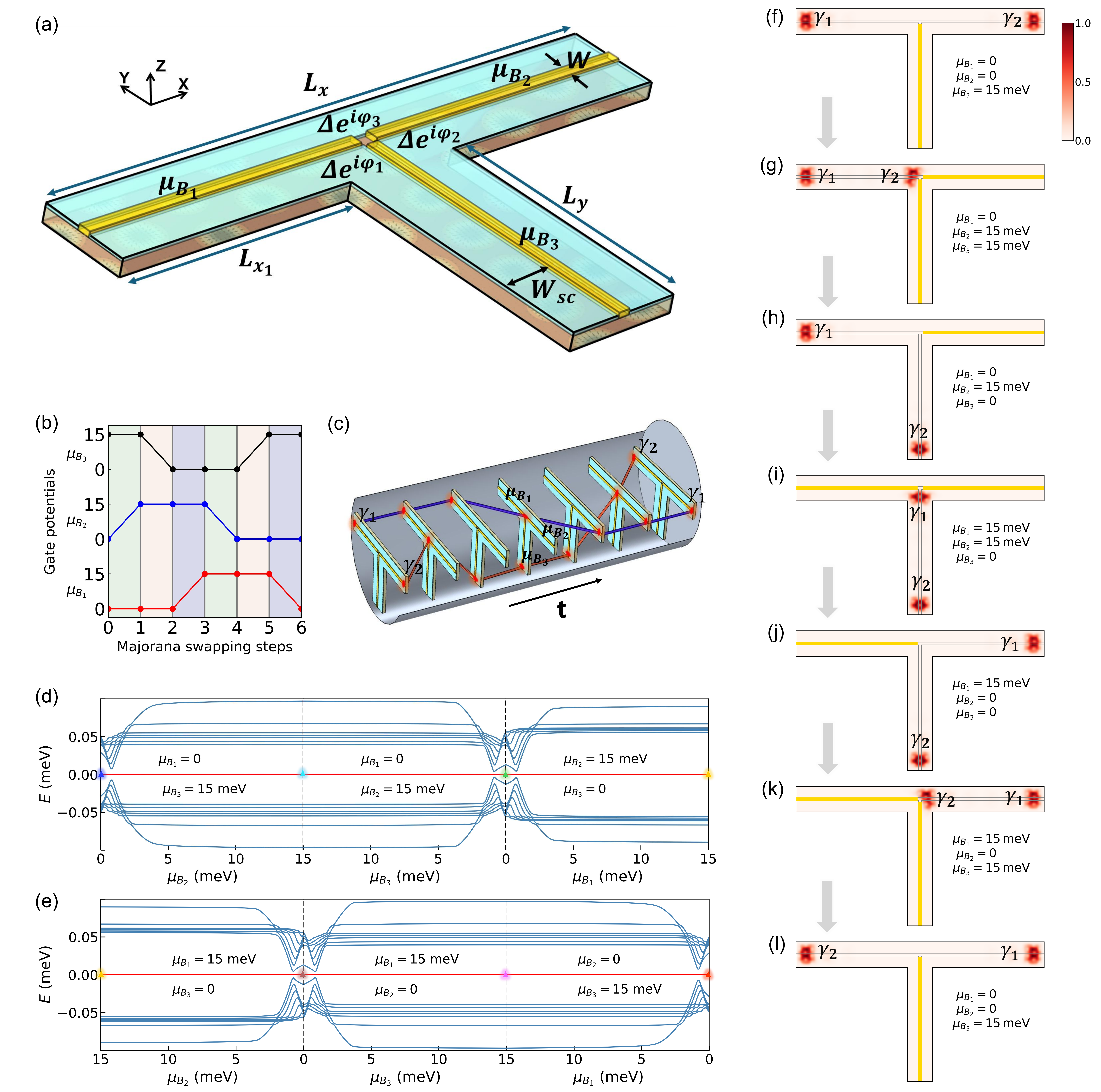,trim=0.0in 0.0in 0.0in 0.0in,clip=false, width=\textwidth}
\caption{(a) A T-shape planar Josephson junction considered for exchanging positions of two Majorana bound states. The gates (in yellow), with tunable chemical potentials $\mu_{B_{1}}$, $\mu_{B_{2}}$ and $\mu_{B_{3}}$, are applied to the non-superconducting channels along the three arms of the device. (b) Steps for changing the values of the chemical potentials. (c) Schematic of the Majorana positions during the exchange. (d)-(e) Variation of the energy eigenvalues with respect to the chemical potentials. (e)-(j) Local density of state (LDOS) of a zero-energy Majorana bound state, at different steps. The dimensions are $L_x$ \!=\! \SI{4.45}{\micro\metre}, $L_{x_1}\!=\! L_y$ \!=\! \SI{2}{\micro\metre}, $W_{sc} \!=\! 200$~nm and $W \!=\! 50$~nm. The length of the gates is \SI{2.18}{\micro\metre}. The global chemical potential is kept at $\mu \!=\! -0.5$~meV.}
\label{FIG:3}\vspace{2mm}
\end{figure*}

\begin{figure*}[t]
\epsfig{file=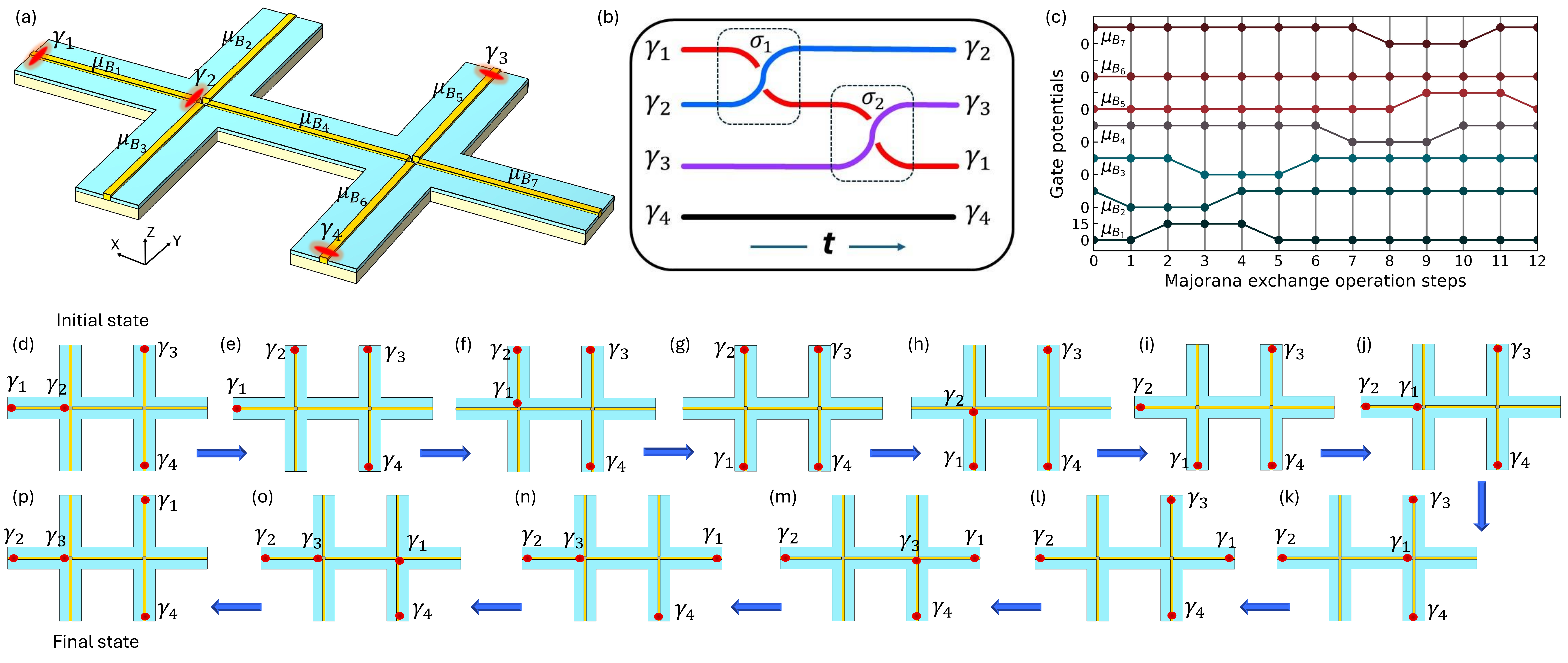,trim=0.0in 0.0in 0.0in 0.0in,clip=false, width=\textwidth}
\caption{(a) A double-cross shaped planar Josephson junction, that can be used for non-Abelian braiding of three Majorana bound states. The external gates (in yellow) can control the chemical potentials $\mu_{B_i}(i=1,\cdot \cdot \cdot, 7)$ in the channels. (b) World line of four Majorana bound states showing two $\sigma$ exchange operations. (c) Gate potential changing steps to perform the $\sigma_1 \sigma_2$ Majorana exchange operation, shown in (b). (d)-(p) Locations of the four Majorana bound states in each step mentioned in (c), to perform a complete $\sigma_1 \sigma_2$ exchange operation.}
\label{FIG:4}\vspace{2mm}
\end{figure*}

\section{Non-trivial fusion of the MBS} 
\label{sec3}
\vspace{-3mm}
The non-trivial fusion rule for two MBS is expressed as $\gamma \times \gamma = {I} + \psi$, where $\gamma$ represents the MBS that come from different MBS pairs, ${I}$ represents the vacuum and $\psi$ represents an unpaired fermionic state~\cite{Beenaker_SciPost2020}. To perform this non-trivial fusion operation, we introduce a barrier gate of length $L_B$ \!=\!\SI{1}{\micro \meter} with a tunable chemical potential $\mu_B$ at the middle metallic channel of our PJJ, as shown in FIG.~\ref{FIG:2}(a). It is incorporated in the Hamiltonian~(\ref{model_hamiltonian}) using an additional term ${\cal H}_{\rm barrier}\!=\!\sum_{i, \sigma} \mu_{B} c^\dagger_{i\sigma} c_{i\sigma}$. A non-zero value of $\mu_B$ effectively splits the PJJ channel into two halves. When $\mu_B$ is large, two pairs of MBS are localized at the two segments of the middle channel. The quasiparticle spectrum as a function of $\mu_B$ is shown in FIG.~\ref{FIG:2}(b). At large positive values of $\mu_B$ ($\mu_B \geq 7$~meV), it is evident that two pairs of zero-energy MBS, protected from higher-energy quasiparticles by an energy gap, robustly appear in the PJJ. The LDOS and CDOS profiles, corresponding to the lowest and second-lowest positive energy eigenstates, at $\mu_B \!=\! 12$~meV, shown in FIGs.~\ref{FIG:2}(c)-(f), reveal two pairs of MBS in the middle channel. This localization characteristic appears in all finite-length PJJs due to hybridization of the wave functions of the two MBS. In Appendix~\ref{appA}, we show the localization of the MBS in a SkX-coupled PJJ of large channel length, in which the true Kitaev limit is realized \textit{i.e.} each zero-energy state from the same MBS pair is localized only at one end of the topological PJJ channel. Nonetheless, since the wave functions corresponding to the two pairs of MBS are isolated in the two segments of the middle metallic channel, as in FIGs.~\ref{FIG:2}(c),(e), for $\mu_B \geq 7$~meV, we denote this regime of $\mu_B$ as the `isolated regime'. In the opposite limit, when $\mu_B$ is small positive ($\mu_B \leq 2$~meV), the two MBS, from different pairs situated near the potential barrier ends, hybridize with each other. When $\mu_B$ is reduced, in this regime, the energy levels of the two MBS pairs move away from zero to higher energies, as shown in FIG.~\ref{FIG:2}(b). The LDOS and CDOS profiles at $\mu_B\!=\!1.1$~meV, in FIG.~\ref{FIG:2}(k)-(n), show the onset of the fusion process, after which only one pair of the MBS remains at the two extreme ends of the PJJ, while the other two near the gate barrier ends disappear. We also uncover that, there exists an intermediate range of $\mu_B$ values between the `isolated regime' and `fusion regime', in which the wave functions of the two MBS pairs are also entangled, as revealed by the LDOS and CDOS profiles in FIG.~\ref{FIG:2}(g)-(j), which show that the MBS wave function, corresponding to one zero-energy eigenstate, is localized at four positions. We denote this regime as `entangled regime' as it shows a non-local quantum entanglement of two different MBS pairs. The `entangled regime' is unavoidable in any finite-size systems, since the MBS wave functions tunnel through the barrier and form a quantum entangled state. We note that the total charge increases by more than an order of magnitude after the non-trivial fusion process takes place, as revealed by FIGs.~\ref{FIG:2}(d),(f),(n), indicative of the creation of a fermionic state after the fusion. The non-trivial fusion of the zero-energy bound states, with two different outcomes (corresponding to the vacuum state and an unpaired fermionic state), can be considered as a hallmark of the non-Abelian character of the MBS.

\section{Trivial exchange of two MBS}
\label{sec4}
\vspace{-2mm}
The exchange of two MBS, belonging to the same pair, can be performed in a T-shape geometry, known to be suitable for this operation~\cite{Alicea_Nature2011}. We use our SkX-coupled PJJ platform in the form of a T junction, as shown in FIG.~\ref{FIG:3}(a). Since the MBS are generated at zero phase difference between the superconducting leads, using three gate potentials $\mu_{B_1}, \mu_{B_2}$ and $\mu_{B_3}$, two MBS can be exchanged, as shown in FIG.~\ref{FIG:3}(b) and (c). FIG.~\ref{FIG:3}(f)-(l) show the LDOS profiles corresponding to the lowest positive-energy eigenstate, describing the different steps during the exchange operation. The quasiparticle energy spectrum as a function of varying gate potentials, shown in FIG.~\ref{FIG:3}(d) and (e), indicates that the two MBS remain at zero energy and protected by an energy gap during this exchange operation. A decrease in the bulk gap is observed when a single gate potential is varied in close to the value of 1 meV, while the other two gate potentials are fixed. During the initial phase of the movement step (\textit{e.g.} step $0 \rightarrow 1$), as the barrier potential $\mu_{B_{2}}$ is changed from 0 to 15 meV, the bulk gap decreases, reaching a minimum at a potential of approximately 1 meV, before subsequently increasing and stabilizing. This modulation induces the translocation of the Majorana bound state $\gamma_2$ from the right terminus to the central region of the T-junction. Furthermore, an analysis of the LDOS within the central region of the T-junction reveals that the maximum LDOS value corresponding to the Majorana bound state becomes equivalent to that of the first bulk state precisely at the $\mu_{B_{2}}$ value where the topological gap is minimized. This observation infers some interaction between the Majorana bound state and the bulk state within this specific range of $\mu_{B_{2}}$. A similar pattern is consistently observed when the other gate potentials are tuned through the critical value of approximately 1 meV during subsequent steps of the exchange procedure. In spite of the reduction of the topological gap during the swapping protocol, the MBS consistently remain at zero energy. This observation also indicates that this mechanism exclusively manipulates the same pair of MBS throughout the exchange procedure.

\section{Non-Abelian braiding of three MBS} 
\label{sec5}
\vspace{-2mm}
Finally, we consider a double-cross geometry that can be constructed using the SkX-coupled PJJ platform, as shown in FIG.~\ref{FIG:4}(a), to perform braiding operation of three MBS. To demonstrate the non-Abelian braiding statistics of the MBS, it is required to show that two subsequent exchange operations involving three MBS, done sequentially in two different ways, lead to different outcomes~\cite{Beenaker_SciPost2020}. This means that the $\sigma_1 \sigma_2$ operation, shown in FIG.~\ref{FIG:4}(b), should yield a different outcome than the $\sigma_2 \sigma_1$ operation. Since the superconducting leads, kept at the same phase, produces topological junctions with MBS at the channel ends, the potential gates placed on top of the channels can be used to move the MBS from one location to another, as shown by the movement scheme in FIG.~\ref{FIG:4}(c) and the physical locations of the MBS in FIG.~\ref{FIG:4}(d)-(p).


\section{Conclusion}
\label{sec6}
\vspace{-2mm}
In this work, we have demonstrated using numerical calculations the non-trivial fusion and braiding operations of zero-energy MBS in a PJJ platform coupled to a SkX. Our device schemes open a practical route to explore the realization and movement of the MBS in the absence of an external magnetic field. This PJJ platform can be flexibly used with both $s$-wave and $d$-wave superconducting leads, expanding the horizon to explore different superconducting materials. Our results establish that the SkX-coupled PJJ platform has the scalability, required for complex device fabrication to perform different logical operations.

The implementation of different quantum gates in the SkX-coupled PJJ platform will be a step forward in the direction of realization of universal quantum computation. For the readout of the initial and final states of the Majorana qubit and quantum gates, fermionic charge or non-Abelian Berry phase can be used as the probes. Despite numerous proposals for platforms in which the braiding, non-trivial fusion and gate operations can be performed~\cite{Sanno_PRB2021,Zhou_NatureComm2022,Bedow2024,Pandey_PRRes2024}, the current magnetic field-free two-dimensional platform with robust gate control brings an unique ability that can be valuable in fault-tolerant topological quantum computation.\\



\section*{Acknowledgements}
\vspace{-2mm}
This work was supported by Science and Engineering Research Board, India (SRG/2023/001188) and an initiation grant (IITR/SRIC/2116/FIG). We acknowledge the National Supercomputing Mission for providing computing resources of PARAM Ganga at the Indian Institute of Technology Roorkee, which is implemented by C-DAC and supported by the Ministry of Electronics and Information Technology and Department of Science and Technology, Government of India. P.S. was supported by Ministry of Education, Government of India via a research fellowship.\\ 

\appendix
\refstepcounter{section}
\setcounter{figure}{0}
\renewcommand{\thefigure}{A\arabic{figure}}
\renewcommand{\theHfigure}{A\arabic{figure}}

\begin{figure*}[t]
\begin{center}
\vspace{-0mm}
\epsfig{file=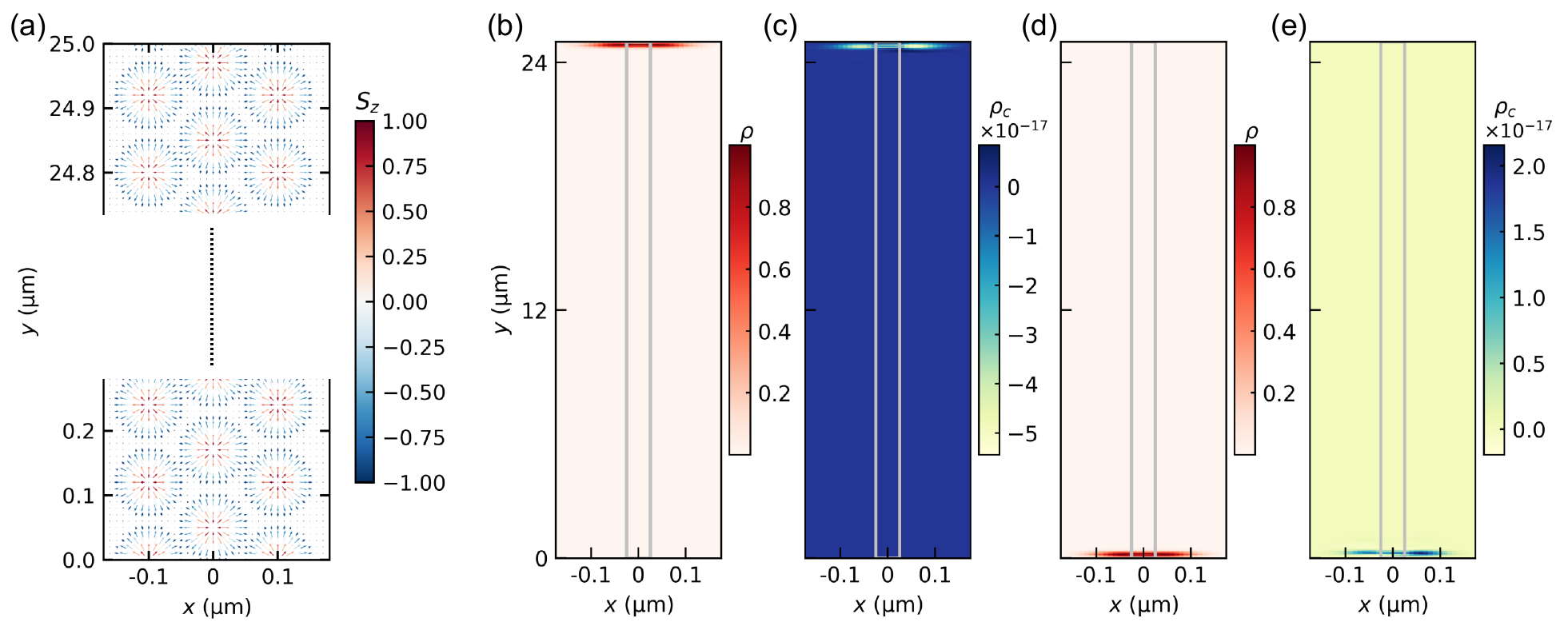,trim=0.0in 0.0in 0.0in 0.0in,clip=false, width= \textwidth}
\caption{(a) Spin configuration of a skyrmion crystal on a lattice of dimensions 350~nm $\times$ \SI{25}{\micro \meter}, used underneath a planar Josephson junction of a very large length. (b), (d) Local density of states (normalized to unity) corresponding to the two lowest positive energy eigenstates. (c), (e) Charge density of states (in units of e/$\upmu$m$^2$) corresponding to the two lowest positive energy eigenstates. The dimensions of the planar Josephson junction and parameters used in these calculations are: $L_x \!=$350~nm, $L_y \!=$\SI{25}{\micro \meter}, $W \!=\! 50$~nm, $g_{eff} \!=\! 40$, $\Delta_{ij} \!=\! 2$~meV $\alpha \!=\! 30$ meV-nm and $\mu \!=\! -0.3$ meV.} 
\label{FIG:A1}
\vspace{-0mm}
\end{center}
\end{figure*}

\section*{Appendix~A: Localization of the MBS in a long PJJ}
\label{appA}
\vspace{-2mm}
In the Kitaev chain model~\cite{Kitaev_2000}, one obtains two MBS that are maximally separated, \textit{i.e.} each Majorana bound state is localized at the opposite ends of the chain. In a PJJ of finite channel length , on the other hand, the localization of a single Majorana bound state occurs at both ends due to hybridization of the wave functions of the two zero-energy MBS, as shown in Fig.~\ref{FIG:1}(d). In a very long PJJ, coupled to a SkX as in Fig.~\ref{FIG:A1}(a), the Kitaev chain limit is restored \textit{i.e.} the MBS pair shows localization only at one end, as shown in Fig.~\ref{FIG:A1}(b)-(e). The PJJs used in experiments are typically shorter than this Kitaev limit(see \textit{e.g.}~\cite{Fornieri_Nature2019,Ren_Nature2019,Banerjee_PRL2023}). Hence, the MBS wave functions will be hybridized, as described in our results presented above. For the above reason, to perform the non-trivial fusion or non-Abelian braiding, it will be useful to maximize the channel length.

\bibliography{Ref}

\end{document}